# Torsional "Superplasticity" of Graphyne Nanotubes

J. M. de Sousa[1], G. Brunetto[1], V. R. Coluci[2], D. S. Galvão[1*]

[1]Applied Physics Department, State University of Campinas, Campinas/SP, 13083-970, Brazil.

[2]School of Technology, State University of Campinas−UNICAMP, Limeira, SP 13484-332, Brazil

**Abstract**

Graphyne is a planar two-dimensional carbon allotrope formed by atoms in $sp$, $sp^2$, and $sp^3$ hybridized states. Topologically graphyne nanotubes (GNTs) can be considered as cylindrically rolled up graphyne sheets, similarly as carbon nanotubes (CNTs) can be considered rolled up graphene sheets. Due to the presence of single, double, and triple bonds, GNTs exhibit porous sidewalls that can be exploited in many diverse applications. In this work, we investigated the mechanical behavior of GNTs under torsional strains through reactive molecular dynamics simulations. Our results show that GNTs are more flexible than CNTs and exhibit "superplasticity", with fracture angles that are up to 35 times higher than the ones reported to CNTs. This GNT "superplastic" behavior can be explained in terms of irreversible reconstruction processes (mainly associated with the triple bonds) that occur during torsional strains.

---

[1]*Corresponding author. Tel +55 (19) 3521-5373. E-mail: galvao@ifi.unicamp.br (Douglas S. Galvao)

1. **Introduction**

The three possible carbon hybridization states ($sp$ and $sp^2$ and $sp^3$) allow different chemical arrangements, which results in a rich variety of allotropic structures in nature [1]. 3D (diamond [2] and graphite [3]), 2D (single layer graphene [4,5]), 1D (nanotubes [6]) and 0D (fullerenes [7]) structures can be formed. These structures exhibit exceptional mechanical and electronic properties [7-10], which have been exploited in a wide range of applications, from superconducting materials to functional nanocomposites [11-13].

Besides graphene, another interesting 2D carbon allotrope is graphyne. Firstly proposed by Baughman *et al.* [14] in 1987, graphyne is a generic name for a family of 2D structures formed by carbon atoms in $sp$ and $sp^2$ and $sp^3$ hybridized states. They exhibit low density, low formation energy, high thermal stability, and interesting electro-mechanical properties [15-20]. Some of graphyne structures have been recently synthesized as thin films on top of copper substrates using cross-coupling reactions [21]. Density functional theory calculations showed that, similarly to graphene, some graphyne structures present Dirac cone features [22-25], with the advantage of having intrinsic non-zero electronic band gaps [26-29]. Besides the finite band gap values, which allow their application in digital electronic devices, graphynes have been also proposed as a material to be employed as selective membranes [30], gas storage [31-34] and lithium ion battery anodes [35].

From a topological point of view, graphyne nanotubes (GNTs) can be considered as rolled up graphyne sheets, in the same way as CNTs can be considered as graphene sheets rolled up into seamless cylinders. GNTs were predicted by Coluci *et al.* in 2003 and similarly to CNTs, armchair, zigzag and chiral types are possible [36,37]. Depending on the parent graphyne sheet, different GNTs can be generated, such as; α-GNT ((Figure 1-b) and γ-GNT (Figure 1-c). GNTs exhibit some interesting electronic properties, for instance, γ-GNT were predicted to exhibit band gap values are that not dependent on tube diameter values or chiralities [38], which is not happen for CNTs.

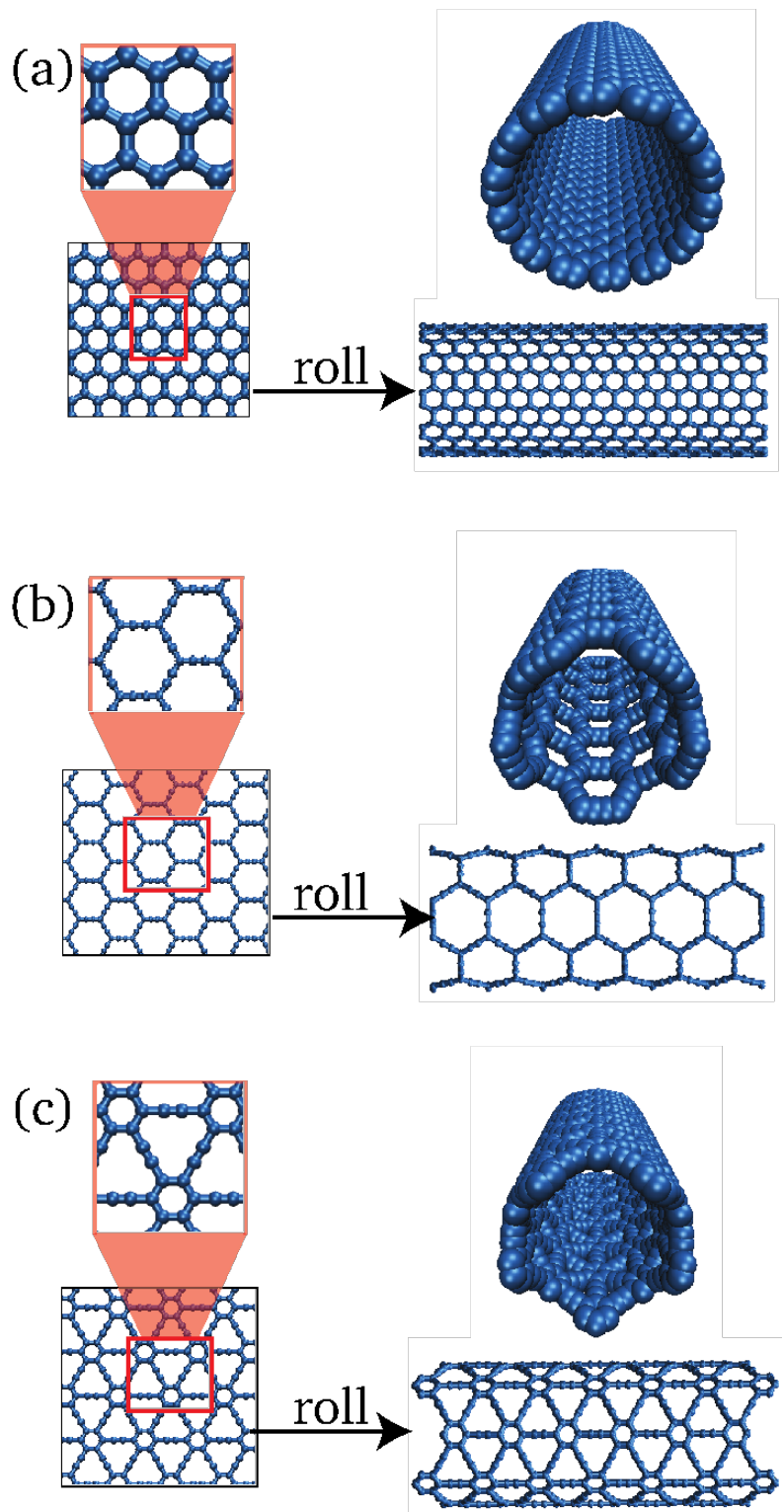

**Figure** 1: Formation of nanotubes (1D) by rolling up 2D sheets into seamless cylinders. The highlighted boxes are a zooming of the different structures showing the atomic arrangement for: (a) graphene; (b) α-graphyne and; (c) γ-graphyne.

Despite the increasing interest in graphynes and GNTs in the last years, to the best of our knowledge, no detailed study of the GNT torsional mechanical properties has been reported. The torsional mechanics of nanotubes under different strain regimes is of particular importance, as it can be the basis for a variety of applications. For instance, CNTs can work as torsional springs when subject to torsional strains. Using atomic force microscopy (AFM), Hall *et al.* [39] were able to determine the CNT torsional spring constant to be 0.41 TPa. The torsional properties of nanotubes have some similarities with the mechanisms of artificial muscles [40] and nanoactuators [41] and their study can be also useful to improve the current models for these structures.

In this work we have carried out an extensive investigation on torsional properties of different GNTs through the use of fully atomistic reactive molecular dynamics simulations.

2. **Methods**

The structures considered in the present work were armchair and zigzag α-GNT (Figure 1-b) and γ-GNT (Figure 1-c), with tube lengths varying from 55 up to 193 Å, and with diameter values from 9 up to 69 Å. These structures are representative of the diverse structural and electronic GNT behaviors [36-38]. For comparison purposes, similar CNT structures (in terms of length, diameter, and chirality values) were also considered.

The torsional tube properties were obtained from the analysis of molecular dynamics simulations carried out using the LAMMPS [42] code, and with the atomistic interactions described by the reactive force field ReaxFF [43]. Charge distributions [44] were calculated based on geometry and connectivity using the electronegativity equalization method [45].

For the analysis of torsional tube loadings, first we divided the GNTs into three main regions (*A*, *B* and *C* - Figure 2). In order to eliminate some pre-stress in the axial direction existing in the initial structure, fully geometric optimization on the entire nanotubes were performed. This optimization was followed by a NVT thermalization of 200 ps at 300 K. The atoms belonging to the region *A* (around the tube axis) were rotated (twisted) with an angular velocity of 0.013 rad/ps until mechanical fracture was observed. The lengths of the fixed regions A and C were 8 Å each. This velocity value is small enough to allow the system to have time to structurally re-equilibrate during the rotation process. Atoms belonging to region *B* were allowed to freely move (thus, the tube length can vary) and they were coupled with a Nosé-Hoover thermostat [46] with a target temperature of 300K (room temperature). No movement along the tube axis direction was allowed for the atoms belonging to regions *A* and *C*. After fixing the atoms belonging to region *A* and *C*, the atoms at region *B* were thermalized by 12.5 ps before starting the rotation/twisting procedures. A time step of 0.25 fs was used in all simulations.

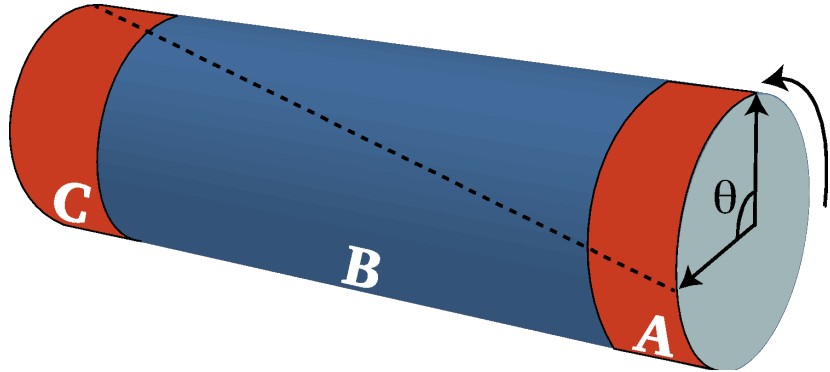

**Figure** 2: Schematic representation of the tubes used in the simulations. Tube extremities, highlighted in red and labeled *A* and *C,* indicate the twisted and the fixed regions, respectively. The central region, labeled *B*, contains atoms that were allowed to freely move. θ refers to the torsional angle value.

The behavior of the GNT under mechanical twisting and the torsional stiffness values were obtained from the analysis of the second derivative of the torsional strain energy as a function of the torsion angle θ, calculated around the equilibrium angles. The energy variation was calculated subtracting the current total energy of the twisted tube from the total energy of the original non-twisted tube.

3. **Results**

In Figure 3 we present representative snapshots from the molecular dynamics simulations of the rotating/twisting processes. As we can see from Figure 3, we observed a collapse of the tube walls just after the half of the first turn (top snapshots in Figure 3). Despite the different visualization angles, which can cause the impression of different patterns, the twisted patterns of the collapsed tubes are the same for the CNT, γ-GNT, and α-GNT. After the collapse and with the increase of the torsion angle value, the structures assume a compact rod-like shape (middle snapshots in Figure 3). CNTs exhibit a rod-like shape at smaller θ values than GNTs. Among GNTs, α-GNT exhibit the rod-like shape at the largest θ values. Continuing to increase the torsion angle causes tubes to break (Figure 3 bottom snapshots). We consider that the tube breaks up when two separate segments can be clearly identified, usually connected by few residual atoms (bottom snapshots in Figure 3). For a better visualization of these processes see movies 01 and 02 in the Supplementary Material.

We have investigated in details how specific tube variables (length, diameter and chirality) affect the critical angle value for fracture. Considering tubes of same chirality, approximately same diameter and length values, we observed an almost linear dependence with tube length. α-GNTs always present the largest fracture angle values, followed by γ-GNTs and CNTs. Results for armchair (diameter ~ 20 Å) and zigzag (diameter ~ 11 Å) tubes are presented in Figure S1 of the Supplementary Material.

For the cases where the rotation/twisting are carried out keeping the tube lengths fixed, there are a significant decrease in the fracture angle values, with an almost exponential behavior as a function of the diameter values, but the resilient ordering to fracture is preserved (see Figure S2 in the Supplementary Material). This kind of exponential behavior was previously also observed for CNTs [44].

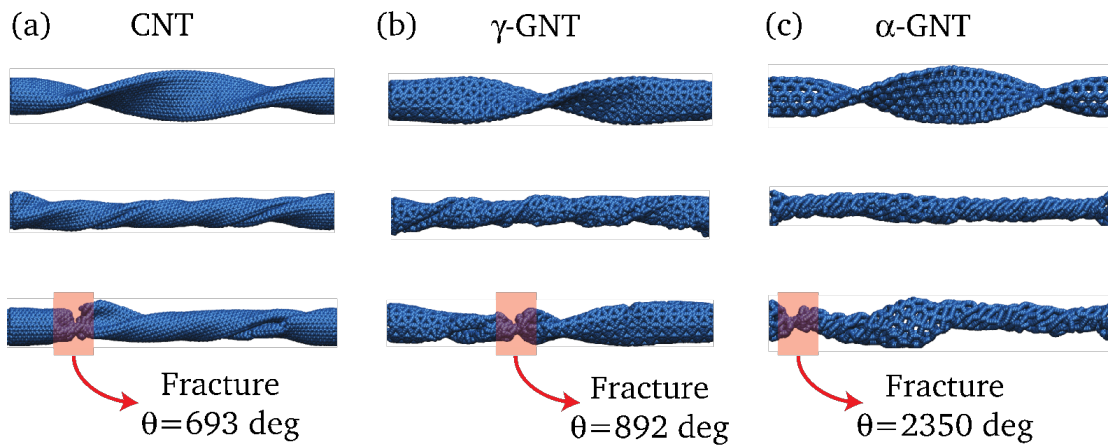

**Figure** 3: Representative snapshots from molecular dynamics simulations of the twisting load for: (a) CNT; (b) γ-GNT; and; (c) α-GNT. Different stages of the twisting processes are shown (from upper to lower): the observed wall tube collapse after one complete turn, followed by configurations where maximum compression is reached just before breaking up, and finally the instant and angle values at which mechanical failure (fracture) were observed. The number of turns needed to reach each stage depends on tube type. The presented results are for the case of armchair tubes of ~180 Å length and ~20 Å diameter values, respectively.

Considering tubes with similar diameters ($d_{armchair}$ = 20Å and $d_{zigzag}$ = 11Å), for armchair tubes the twist stiffness for CNT is up to 10.4 times higher than α-GNT and 2.6 times higher than γ-GNT, while these numbers change to 35.0 and 4.8, respectively, for the zigzag tubes. In general, the stiffness decreases exponentially with the increase of the tube length (keeping constant the diameter).

On the other hand, torsional stiffness exponentially increases as diameter increase (and fixed tube length). The torsional stiffness for CNT is up to 1.6 (2.4) and 7.6 (35.3) times higher than armchair (zigzag) γ-GNT and α-GNT, respectively (see Supplementary Material).

Our results also indicated that it is necessary more energy to twist and also to break CNTs, in comparison with γ-GNTs and α-GNTs (Figure 4). As observed by Wang et al. [47], the accumulated torsional energy in the case of CNTs increases continuously until the point where an abrupt drop energy value (tube fracture) is observed. On the other hand, GNT exhibit a sequence of small energy drops before fracture. Whereas γ-GNT presents only one or two small energy drops before fracture, α-GNT presents around 10 energy drops (Figure 4). Similar behaviors were observed for both armchair and zigzag tubes. The small energy drops come from successive structural irreversible reconstructions (Figure 5 for γ-GNT and Figure 6 for α-GNT). The reconstructions are induced by tensions generated during torsional loads and occur only for GNTs. For CNTs, the carbon atoms are organized in compact honeycomb geometry, with not enough free space to allow extensive atomic rearrangements. These compact configurations lead to relative fast fractures. As GNTs are in general more porous than CNTs, there is more free space to allow atom moves and rearrangements before the structural break up.

These complex reconstruction processes are schematically explained using Figure 5, for the case of a γ-GNT. In this Figure the main tube axis is aligned with the red line. The red line also indicates the rotation axis and the arrows the rotation directions. The yellow ring highlights the sites where reconstruction occur. Due to the torsion load, the tube initially changes to a compact rod-like stage (as discussed previously). The simulations also revealed that the tensile stress is higher along the direction that crosses the highlighted ring, as shown in Figure 5-b. This tension causes ring elongations (Figure 5-b), increases the bond length values of the bonds parallel to the tension direction, and creates space to incorporate one of the neighboring atoms (blue atoms - Figure 5-b). The neighboring atom is incorporated creating a metastable heptagon ring (Figure 5-c). For the cases considered here we only observed 1-2 hexagon → heptagon reconstructions before tube fractures. The fracture usually starts from one of the defect sites (heptagon rings) generated by the reconstruction. This was the reason to observe only one or two energy drops in the γ-GNT energy profile before breaking up.

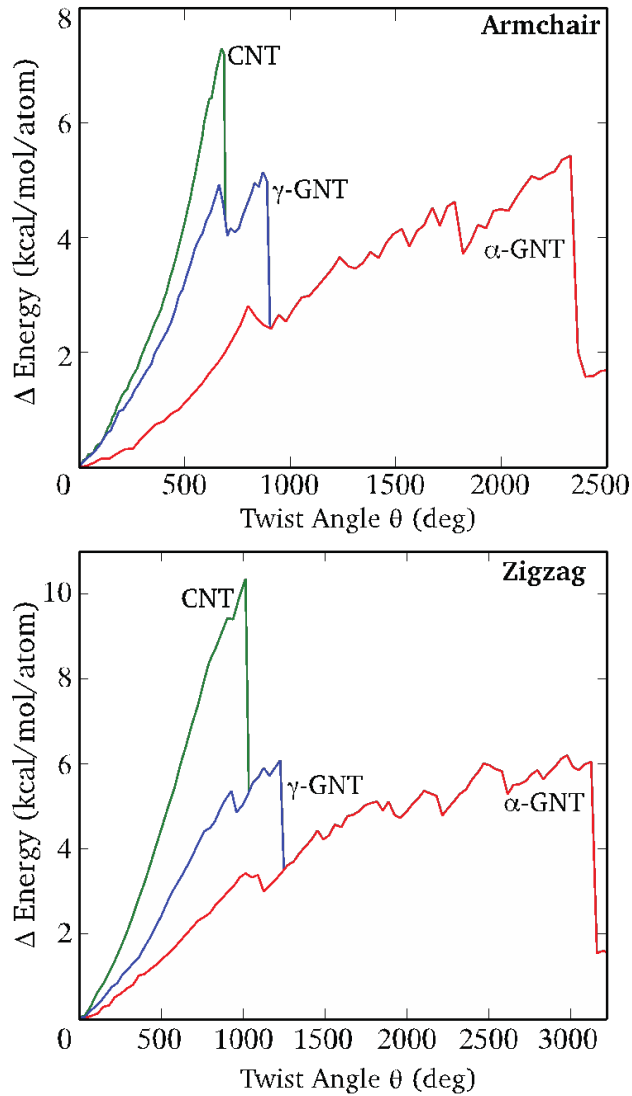

**Figure** 4: Energy variation (ΔE) during the twist loading processes. Carbon nanotubes (green line) require more energy to be twisted in comparison with both α- and γ-GNTs. However, CNTs in general present only one energy drop, which occurs when the tubes break up. γ-GNT presents around one or two small energy

drops before the structure beaks up. α-GNT presents many small energy drops (around 10) before break up. These small energy drops are related with successively structure reconstruction that is associated with the α-GNT super-elastic behavior. All considered tubes have lengths ~192 Å.

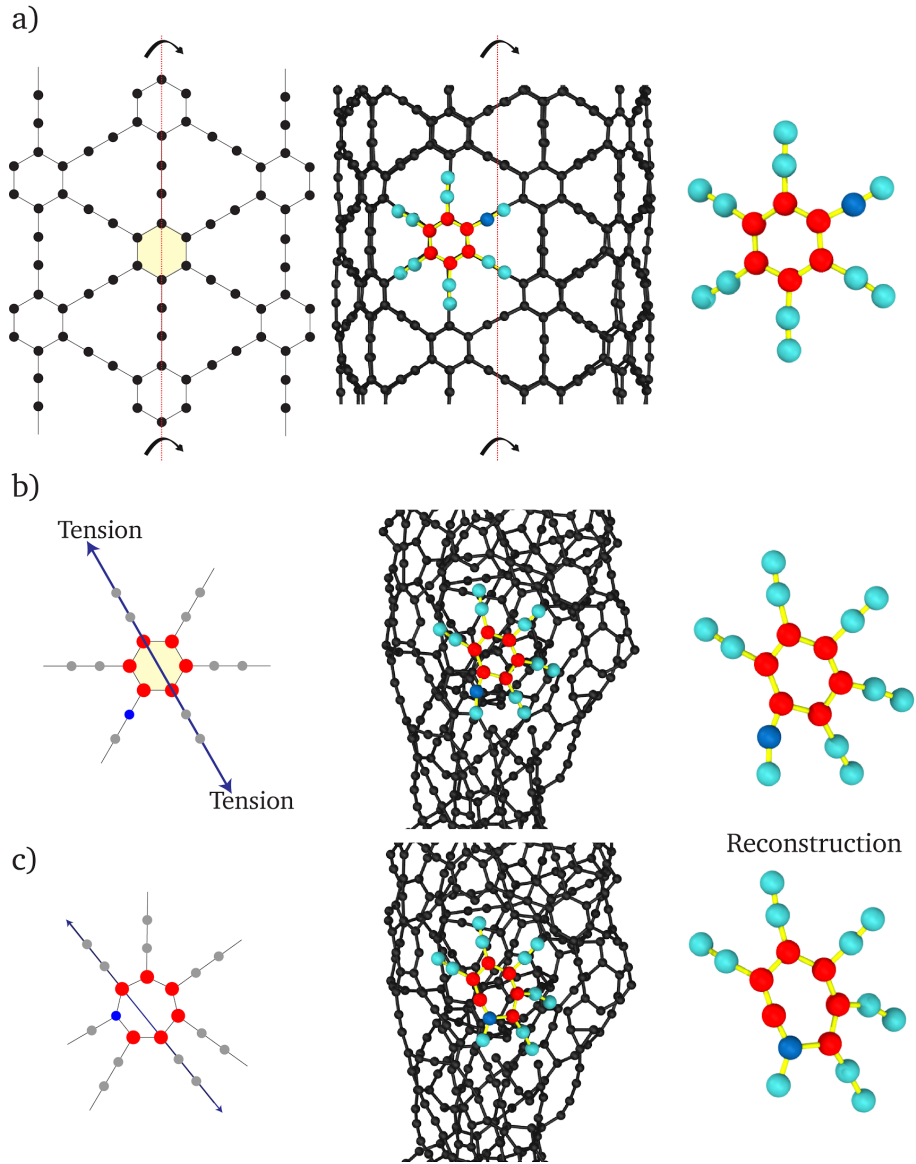

**Figure** 5: Scheme (left) and simulation snapshots (center and right) of the reconstruction process during the twist of a γ-GNT. (a) γ-graphyne structure. The tube main axis in aligned along the red line. The arrows represent the twist directions. The highlight ring indicates the sites where the reconstruction occurs. (b)

Direction of the tension generated due to the twist load. (c) Reconstruction of the structure after incorporating some (blue) atoms. This reconstruction creates a fragile point where the fracture originates.

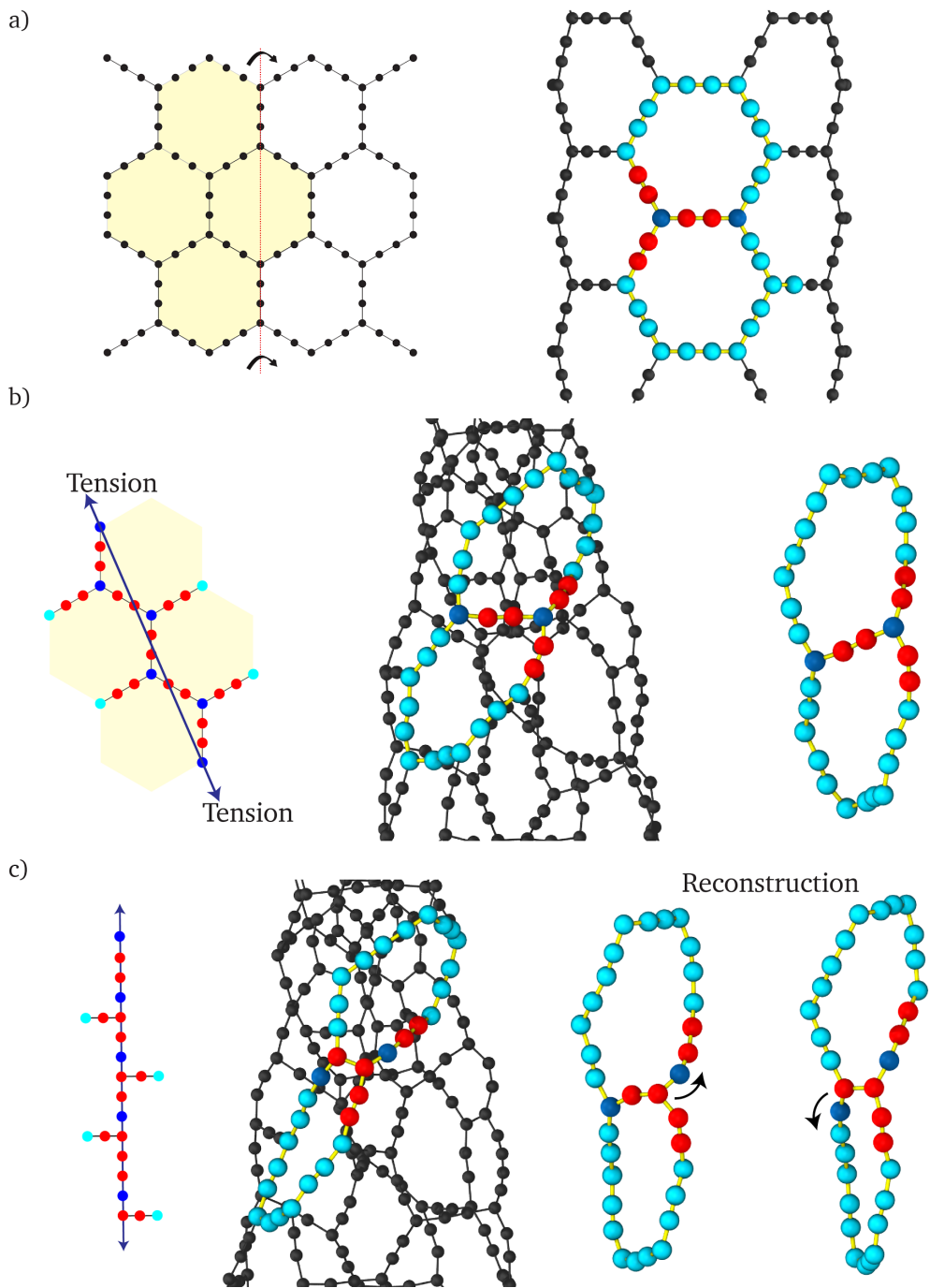

**Figure** 6: Scheme (left) and simulation snapshots (center and right) of the reconstruction process during the twist of a α-GNT. (a) α-graphyne structure. The tube main axis in aligned along the red line. The arrows represent the twist directions. The highlight rings indicate the sites where reconstruction occurs. (b) The blue atoms left the position of join two rings and are fused into the linear atomic carbon chain (in the direction of the tension line). (c) Once the incorporated atoms occurs in the line where the tension arises, there is an instantaneously tension released due to an increase of the chain length.

Reconstructions also occurred for α-GNT but with more event occurrences than for γ-GNT. During the torsion load for α-GNT (Figure 6-a), the tubes are stressed along a line that crosses many rings (Figure 6-b). The carbon atoms are pulled off in the direction of the created tension. The stress accumulated on the atomic carbon chain (in the direction of the tension line - Figure 6) is released by incorporating some carbon atoms (in blue) that are already connected to the chain, increasing the atomic chain length.

This very efficient mechanism of releasing tensile stress through the incorporation of neighbor carbon atoms allows the α-GNT to stand for a larger number of turns before breaking, in comparison with the γ-GNT and CNT. This reconstruction mechanism is in the origin of α-GNT superplasticity behavior during torsion loads. The fact that α-GNTs can stand larger twist loads (superplastic-like behavior) involves irreversible structural reconstructions, which make α-GNTs unable to recover their original geometry after the completion of unloading processes (see Supplementary Material).

The reconstruction events for armchair and zigzag α-GNT can be identified through the small energy drops in the energy profiles (Figure 4, around 800 and 2200 degrees for armchair and 1000 and 3000 degrees for zigzag, respectively). The reconstruction occurs while carbon atoms (in blue - Figure 6) are incorporated into the linear carbon atom chain (Figure 6). When the reconstruction process stops, some regions become structurally fragile (red and light blue atoms - Figure 6) and break.

4. **Summary and Conclusions**

We studied the mechanical properties of graphyne nanotubes (GNTs) under torsion loads by fully atomistic reactive molecular dynamics (MD) simulations using the ReaxFF potential. We have considered α-GNTs and γ-GNTs of different lengths, diameters and chiralities (armchair and zigzag). Carbon nanotubes (CNTs) of similar dimensions were also considered for comparative purposes.

Our results indicated that GNTs are more resilient than carbon nanotubes (CNTs) to torsional fractures. They also have higher torsional stiffness that increases exponentially as the tube diameter increases. Also,

GNT were predicted to exhibit a superplasticity behavior and they can break at angles up to 3 times larger than the ones predicted for CNTs. From the MD simulations was possible to identify the precise origin of the superplasticity behavior. Our results show that it is a direct consequence of multiple irreversible reconstructions (mainly associated with the triple bonds) during the torsional process with the incorporation of atoms from the ring structures into linear atomic chains.

5. **Acknowledgements**

This work was supported in part by the Brazilian Agencies CAPES, CNPq and FAPESP. The authors thank the Center for Computational Engineering and Sciences at Unicamp for financial support through the FAPESP/CEPID Grant # 2013/08293-7.